\documentclass[%
superscriptaddress,
 amsmath,amssymb,
 aps,
 prl,
 reprint,
floatfix,
]{revtex4-2}

\usepackage{graphicx}
\usepackage{dcolumn}
\usepackage{bm}

\usepackage{blindtext}

\begin{document}

\preprint{APS/123-QED}

\title{History effects in the creep of a disordered brittle material}

\author{Tero M\"{a}kinen}
 \email{Corresponding author\\tero.j.makinen@aalto.fi}
\affiliation{%
 Department of Applied Physics, Aalto University,\\
 P.O.  Box 11100, 00076 Aalto, Espoo, Finland
}%
\affiliation{NOMATEN Centre of Excellence, National Centre for Nuclear Research, ul. A. Soltana 7, 05-400  Otwock-\'{S}wierk, Poland}
\author{J\'er\^ome Weiss}
\affiliation{%
 Univ. Grenoble Alpes, CNRS, ISTerre, 38000 Grenoble, France
}%
\author{David Amitrano}
\affiliation{%
 Univ. Grenoble Alpes, CNRS, ISTerre, 38000 Grenoble, France
}%
\author{Philippe Roux}%
\affiliation{%
 Univ. Grenoble Alpes, CNRS, ISTerre, 38000 Grenoble, France
}%


\date{\today}

\begin{abstract}
We study the creep behavior of a disordered brittle material (concrete) under successive loading steps, using acoustic emission and ultrasonic sensing to track internal damage. The primary creep rate is observed to follow a (Omori-type) power-law decay in the strain rate, the number of acoustic emission events as well as the amplitudes of the ultrasonic beams, supporting a brittle-creep mechanism. The main outcome is however the discovery of unexpected history effects that make the material less prone to creep when it has been previously deformed and damaged under primary creep at a lower applied stress. With the help of a progressive damage model implementing thermal activation, we interpret this as an aging-under-stress phenomenon: during an initial creep step at relatively low applied stress, the easy-to-damage sites are exhausted first, depleting the excitation spectrum at low stress gap values. Consequently, upon reloading under a larger applied stress, although previously damaged, the material creeps (and damages) less than it would under the same stress but without pre-creeping. Besides shedding a new light on the fundamental physics of creep of disordered brittle materials, this has important practical consequences in the interpretation of some experimental procedures, such as stress-stepping experiments. 

\end{abstract}

\pacs{Valid PACS appear here}
\maketitle

The failure of materials is naturally important for practical applications but also for the fundamental physics involved. In everyday applications, such as in structural materials, the loads are often static, leading to time-dependent deformation  -- creep -- and possibly failure.

Creep behavior of materials is usually divided into distinct phases. The first phase, primary creep, is characterized by a power-law decrease in the strain rate $\dot{\epsilon} \propto t^{-p}$ where the exponent $p$ is reported to range between ~0.4 and 1 (e.g. \cite{carter1978transient}). Andrade historically reported an exponent $p=2/3$ for metal wires~\cite{andrade1910viscous}, defining the so-called Andrade's creep, while $p=1$ defines logarithmic creep (as strain increases as $\log(t)$)~\cite{cottrell1997logarithmic}. After primary creep one might enter secondary creep, where the strain rate reaches its minimum value $\dot{\epsilon}_m$ and stays constant. Finally, the material fails due to a rapid increase in the strain rate, called tertiary creep~\cite{nechad2005creep,leocmach2014creep,koivisto2016predicting,makinen2020scale}. A classical goal is to determine a creep law of the form $\dot{\epsilon}_m=A\sigma^n \text{exp}(-E/k_BT)$, where the exponent $n$ is assumed to be linked to a underlying microscopic mechanism, $T$ is the temperature, $k_B$ the Boltzmann's constant, $E$ a stress-independent but material-dependent activation energy and $A$ a material constant \cite{ashby2012engineering}. In this empirical law, the Arrhenius term expresses the thermally-activated nature of creep. 

In practice, a stress-stepping procedure is often used for creep testing~\cite{heap2009time, brantut2013time, geng2018time} where the sample is first loaded to a stress level $\sigma$ for a period of time to study the creep rate and then successively loaded to higher stress levels. This can be seen as a cost-effective testing method and also as a way to eliminate the sample-to-sample variation in heterogeneous materials (as multiple stresses can be tested on the same specimen, at a given temperature) in order to determine the creep law of the material, assuming that the minimum strain-rate $\dot{\epsilon}_m$ is attained at the end of each step.
However a question naturally arises: are there history effects in this process, i.e. are the previous creep stages affecting the current one?
In the context of dislocation-driven creep in metals, such a possibility was already hinted by Cottrell~\cite{cottrell1952time} who argued that the creep rate in the latter creep stages would be slower as some of the easy-to-deform sites have already been "exhausted" in the initial creep stages. In this Letter we show, 
from experimental observations of the global strain, acoustic emission and ultrasonic sensing on an emblematic quasi-brittle heterogeneous material -- concrete -- that creep deformation and damage are indeed characterized by history effects leading to unexpected relations between the applied stress and the strain-rate. By simulating a simple progressive damage model, we show that this effect is due to an exhaustion mechanism -- a form of \emph{aging-under-stress} -- that can make a material, previously deformed and damaged during a primary creep stage, less prone to further deform later under a \emph{larger} stress.
These results shed a new light on the fundamental physics of creep, and might challenge the interpretation of some experimental results, e.g. those based on stress-stepping. 

The phenomenological power-law decrease of $\dot{\epsilon}$ during primary creep has been observed for a wide range of materials, including metals~\cite{andrade1910viscous}, paper~\cite{koivisto2016predicting}, colloidal glasses~\cite{siebenburger2012creep}, gels~\cite{leocmach2014creep}, ice~\cite{glen1955creep} or rocks~\cite{scholz1968mechanism, heap2009time}.
In various crystalline materials, primary creep mechanisms have been discussed in terms of dislocation interactions~\cite{mott1953lxxviii,cottrell2004microscopic,louchet2009andrade, miguel2002dislocation}. In rocks, it has been proposed that primary creep results from the cumulative effect of microfracturing/damage events, defining the so-called brittle creep~\cite{scholz1968mechanism}.  
We have chosen concrete for creep testing, a disordered material of obvious interest in civil engineering.
The microscopic origins of concrete creep are still partly unknown \cite{benboudjema2005interaction} and are usually understood to relate e.g. to the rearrangements of calcium-silicate-hydrates (primary component of the cement paste)~\cite{vandamme2009nanogranular} or to the migration of adsorbed water in the micro- and nanoporosities \cite{benboudjema2005interaction}, and modelled using viscoelastic approaches~\cite{bavzant1998fracture,briffaut2012concrete}.
Here we however take another viewpoint, and later confirm it, of brittle creep~\cite{scholz1968mechanism} where creep results from microcracking. These microcracking events can be indirectly observed as crackling noise~\cite{sethna2001crackling}.
A popular method for observing this crackling noise is Acoustic Emission (AE) monitoring, which has been used in detecting avalanches in compression of a wide variety of materials e.g. porous brittle materials~\cite{baro2013statistical}, rocks~\cite{lockner1993role,davidsen2007scaling}, solid foams~\cite{makinen2020crossover,reichler2021scalable}, and wood~\cite{makinen2015avalanches}.
Microcracking avalanches have been studied in monotonic loading of concrete using AE~\cite{vu2019compressive, vu2020asymmetric} and the avalanche energy statistics have shown a scalefree power-law distribution with an upper cutoff, $p(E_{AE}) \propto E_{AE}^{-\beta} \exp\left( - E_{AE}/E_0 \right)$. The cutoff $E_0$ diverges as the sample approaches the critical failure stress and this critical transition has been mapped to the universality class of depinning~\cite{fisher1998collective, dahmen2009micromechanical, narayan1993threshold}. We show here that the damage avalanche statistics during primary brittle creep are instead characterized by robust power-law distributions without a cutoff, shining light on fundamental differences between athermal (under monotonic loading) and creep (thermally activated) deformation of disordered materials.

{\em Experiments -} The material studied was lab-size samples of concrete with two different microstructures, one with medium size aggregates (denoted by M) and one with coarse aggregates (denoted by C). For details see~\cite{SM}.

To study the effect of increasing load on the same sample, the creep loading was done under uniaxial compression in successive creep steps at room temperature. This means doing a steep ramp of constant stress rate to a stress level $\sigma_1$, keeping the stress constant for time $T_1$, then doing a second ramp to stress $\sigma_2$, keeping the stress constant for time $T_2$ and so forth. This was done until sample failure which, for the experiments presented here, happened during one of the ramps, giving also the failure stress of the sample $\sigma_c$.

Additionally the crackling noise or Acoustic Emission (AE) in the sample was monitored using two piezoelectric sensors, which yields a catalog of acoustic event energies $E_{AE}$. 
The sample was simultaneously monitored using ultrasonic tomography~\cite{tudisco2015timelapse} using two arrays of ultrasonic transducers in order to follow the evolution of the attenuation of direct (also called ballistic) elastic waves within the material, a way to probe internal damage of the sample.
See \cite{SM} for full details on the experimental methods.

\begin{figure}[t!]
\includegraphics[width=\columnwidth]{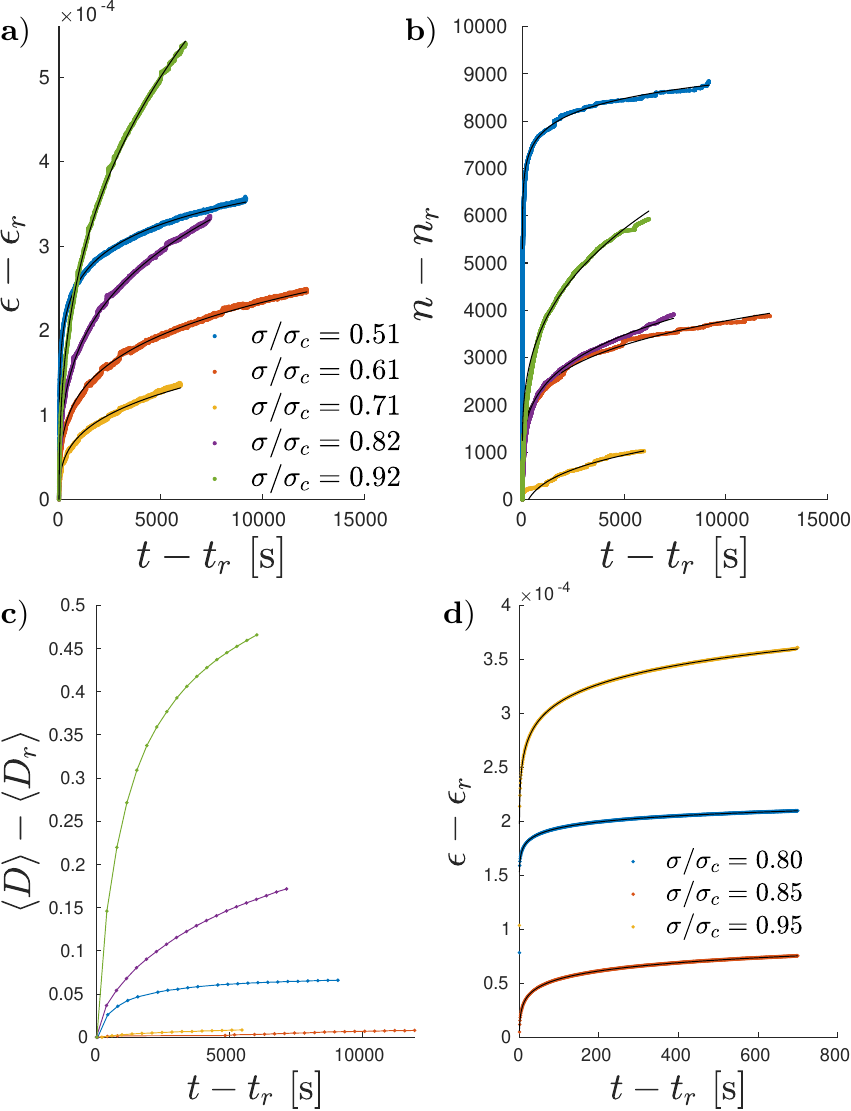}
\caption{\label{fig:fig1} a) Strain $\epsilon$ accumulated in the sample after the start of each creep stage as a function of the time $t$ after the start of the creep stage at $t_r$. 
b) Same as (a) for the number of acoustic events $n$ accumulated after the start of each creep stage. 
c) Same as (a) for the damage parameter $D$, averaged over all the beams.
d) Same as (a) for the simulations of the progressive damage model with thermal activation. The black solid lines represent the fits according to Eq.~\ref{eq:omoriStrain}.}
\end{figure}

{\em Model -} To explore the behavior computationally, we used a progressive damage model, which the athermal version has been extensively detailed elsewhere ~\cite{amitrano1999diffuse, girard2010failure, girard2012damage}. Here the time-dependent, thermally-activated creep process is introduced using a Kinetic Monte Carlo~(KMC) algorithm~\cite{voter2007introduction} to allow the damaging of element~$i$ with a rate $\nu_i(\sigma,T)=\nu_0 \exp(-E_i/k_BT)$, where $\nu_0$ is an attempt frequency, and the local activation energy writes $E_i=V_a\Delta\sigma_i$ with $V_a$ a constant activation volume and $\Delta\sigma_i$ the Coulomb stress gap between the local stress state and the failure envelope.
The simulation protocol used here, under uniaxial compression, corresponds to initially loading the sample to 80 \% of the maximum stress (determined from monotonically and athermally loading the same microstructure until failure) for 700~s and then increasing the stress to either 85~\% or 95~\% of the maximum stress for additional 700~s. The simulations are averaged over 100 realizations of the microstructure.
For full details on the simulations, see~\cite{SM}.\\

{\em Results -} Here for the sake of clarity, we have focused on a single representative experiment (see \cite{SM} for results from additional experiments).
Looking at the accumulated strain from the beginning of each creep step (see Fig.~\ref{fig:fig1}a) we observe the strain rate decreasing with time. By assuming an Andrade-like law~\cite{andrade1910viscous} to hold and avoiding the divergence at the start of the creep step where~$t=t_r$ (by introducing a time constant $c$) one arrives at the equation
\begin{equation} \label{eq:omoriStrain}
    \dot{\epsilon} = \frac{K}{(t-t_r+c)^p}
\end{equation}
where $\dot{\epsilon}$ is the strain rate, $t-t_r$ the time from the start of the creep at $t_r$, $p$ an exponent and $K$ a prefactor. The time constant was found to be small compared to the duration of creep steps, and not to vary significantly between different steps and experiments, so a constant $c = 1$~s was set.
By integrating Eq.~\ref{eq:omoriStrain} and fitting to the strain data (see black lines in Fig.~\ref{fig:fig1}a) we see that it works extremely well. We observe a slight decrease in the value of the exponent $p$ with increasing applied stress~\cite{SM}, consistently with former observations \cite{carter1978transient, cottrell1952time} and recent simulations~\cite{weiss2022logarithmic}.

On the same fits one would expect the prefactor $K$ to have a power-law type of dependency on the applied stress $K \propto \sigma^m$~\cite{carter1978transient}, or at least a monotonically increasing one.
This is complicated in our case  by the slight change in the $p$-value, nevertheless one would expect the creep rate to increase with increasing stress. This is clearly not the case in our experiments (Fig.~\ref{fig:fig1}a): after the initial creep stage at $0.51 \times \sigma_c$, primary creep is comparatively slower during the following stress steps ($0.61 \times \sigma_c$ to $0.82 \times \sigma_c$)  -- until one reaches the stress of $0.92 \times \sigma_c$ at which creep dynamics finally surpasses that of the initial stress step.

\begin{figure}[t!]
\includegraphics[width=\columnwidth]{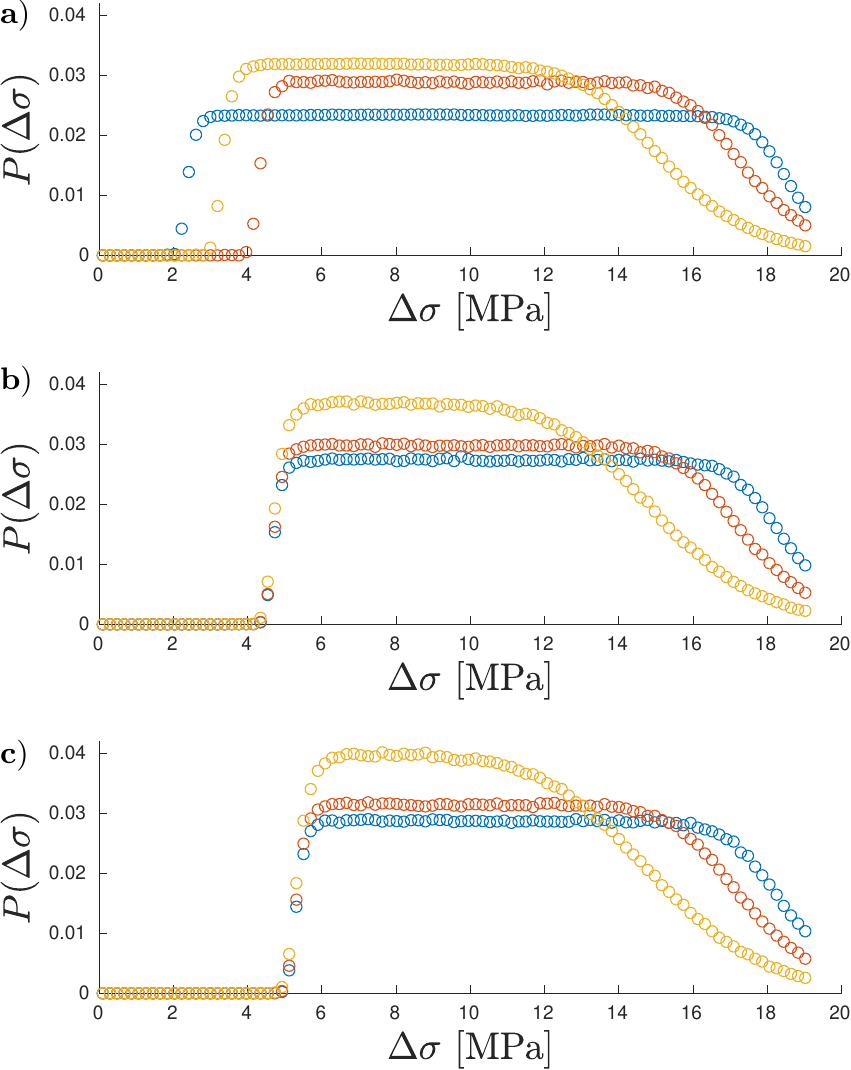}
\caption{\label{fig:fig2} 
The distribution of the Coulomb stress gaps $\Delta \sigma$ at a)~$t-t_r = 0.35$~s b)~$t-t_r=10$~s and c)~$t-t_r=700$~s in the simulations. Blue dots correspond to the initial step at $\sigma = 0.80 \times \sigma_c$, red dots to the step at $\sigma = 0.85 \times \sigma_c$, and yellow ones to $\sigma = 0.95 \times \sigma_c$.}
\end{figure}

If creep results from the cumulative effect of microcracking events~\cite{scholz1968mechanism}, one would expect the strain rate to be directly proportional to the rate of the number of acoustic events $\dot{n}$, which  would have therefore an Omori-type power-law decrease of the form
\begin{equation} \label{eq:omoriEvent}
    \dot{n} = \frac{K^\prime}{(t-t_r+c)^p}
\end{equation}
where $K^\prime$ is a different prefactor. Indeed we see this clear decrease in the event rate in the experimental data (see Fig.~\ref{fig:fig1}b). To check the proportionality $\dot{\epsilon} \propto \dot{n}$, which means $K \propto K^\prime$, we take the fits from Fig.~\ref{fig:fig1}a and apply a linear transformation to yield fits for the number of acoustic events (black lines on Fig.~\ref{fig:fig1}b). The agreement observed strongly argues for the brittle creep hypothesis for concrete. It also implicitly suggests that the shape of the distribution of microcrack sizes, i.e. of AE energies, does not evolve during primary creep as well as from one loading step to another (see below).

We also measured the amplitude of the direct wave between each pair of source-receiver ultrasonic transducers $A$, normalized by their amplitude at the beginning of the first creep step $A_0$, and argue that a decrease in this amplitude reflects a damaging process. The damage parameter, averaged over all the beams, is therefore defined as
\begin{equation}
    \langle D\rangle = \left\langle \frac{A_0 - A}{A_0} \right\rangle
\end{equation}
The evolution of this damage parameter (see Fig.~\ref{fig:fig1}c) is very similar to the evolution of the strain or the number of events: it grows with time with a decreasing rate, and in the stress steps directly after the initial $0.51 \times \sigma_c$ one, the damage accumulation is slower. This confirms the brittle creep mechanism as well as the associated history effects.

The simulation results from our progressive damage model agree remarkably well with these experimental findings: the strains follow Eq.~\ref{eq:omoriStrain} (black lines in Fig.~\ref{fig:fig1}d), and the history effect on the creep rate is evident when the stress of the second creep stage is small enough ($\sigma = 0.85 \times \sigma_c$) while with a high enough stress ($\sigma = 0.95 \times \sigma_c$) the initial rate (at $\sigma = 0.80 \times \sigma_c$) is surpassed, as observed on Fig.~\ref{fig:fig1}a.
Similarly to the experiments, a decrease in the exponent~$p$ of Eq.~\ref{eq:omoriStrain} is seen with increasing stress (see \cite{SM} for details).\\

The model allows to ascribe these history effects on creep deformation to a mechanism of aging-under-stress. We do this by examining the excitation spectra~\cite{lin2014density, lin2014scaling, liu2016driving, ovaska2017excitation}, i.e. in our case the Coulomb stress gap distribution (see \cite{SM} for a full derivation). The simulations start with a uniform cohesion distribution mimicking microstructural disorder, leading to a uniform stress gap distribution. However, this distribution is rapidly depleted towards the small stress gaps as primary creep proceeds (Fig.~\ref{fig:fig2}), as the easy-to-damage sites are exhausted. This leads to a slowing down of the creep rate in the second step with $\sigma = 0.85 \times \sigma_c$, for which the lower cut-off of the stress gap distribution does not almost evolve upon reloading. Increasing further the applied stress to $\sigma = 0.95 \times \sigma_c$ counterbalances this aging effect and the stress gap distribution significantly narrows at both ends during creep. Coupled with a general softening of the material as damage accumulates, this makes the creep dynamics to finally surpass the one of the initial stress step.

\begin{figure}[t!]
\includegraphics[width=\columnwidth]{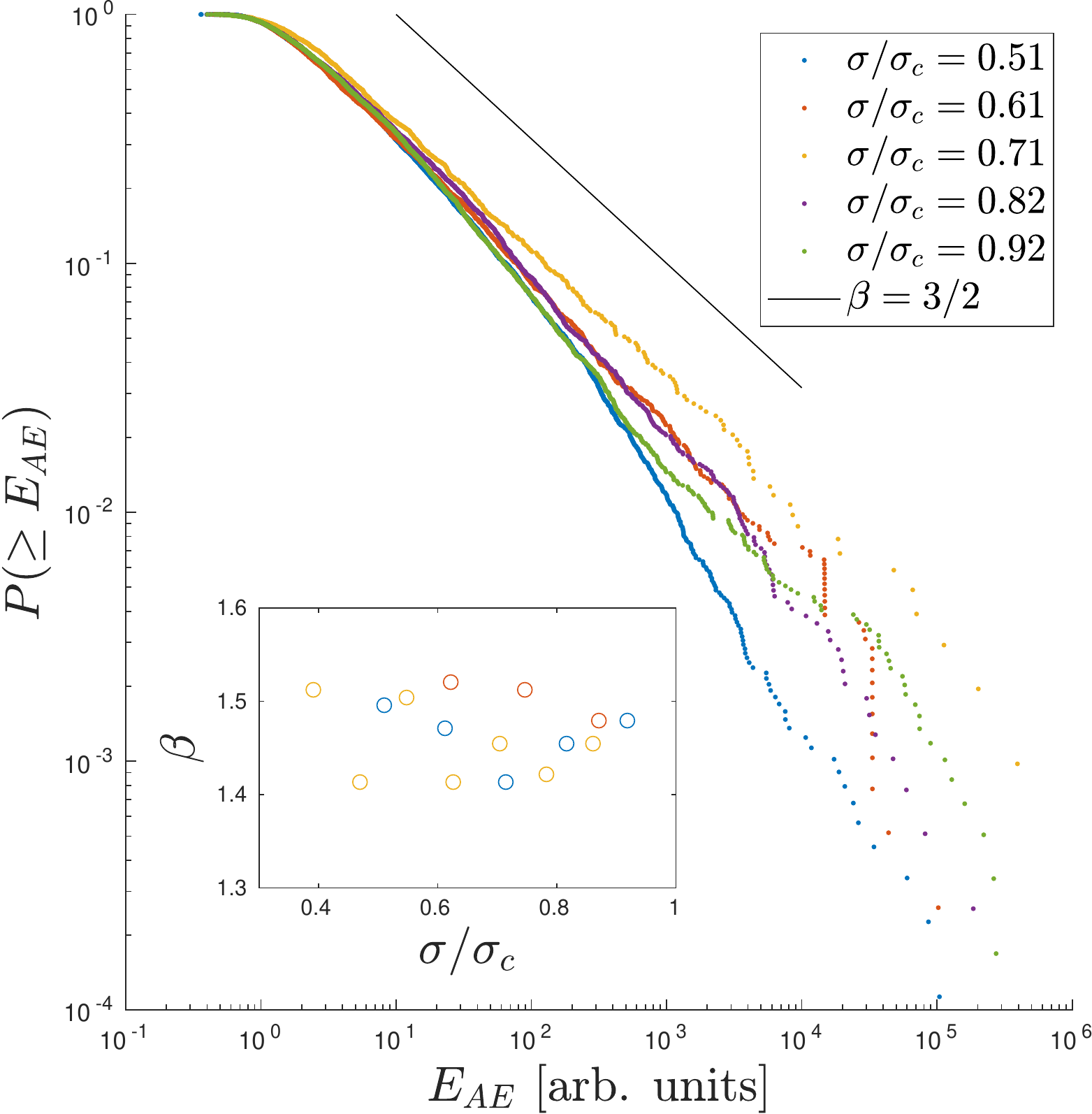}
\caption{\label{fig:fig3} The cumulative distribution of acoustic emission event energies $E_{AE}$ for each of the creep steps in one experiment. The inset shows the values of the power-law exponent $\beta$ obtained through a maximum likelihood estimation for all the stress steps.}
\end{figure}

In addition to the number of acoustic events, one can take a look at the distribution of their energies. Plotting the cumulative distribution of energies $P(\geq E_{AE})$ (see Fig.~\ref{fig:fig3}) one can see a power-law type distribution spanning several decades in energy. 
As already mentioned, for (athermal) monotonic loading of concrete, the distribution of event energies was found to follow a truncated power-law where the cutoff scale $E_0$ diverges as the sample approaches failure, signing a divergence of the correlation length and a critical transition ~\cite{vu2019compressive}. Creep fundamentally differs on this respect, as we do not observe such a cutoff, even under stresses far from the monotonic failure stress $\sigma_c$ of the material and/or during primary creep, i.e. far from the creep failure time.  Instead the distributions follow a pure power-law $p(E_{AE}) \propto E_{AE}^{-\beta}$. 
Note that a truncated power law cannot be statistically excluded (see \cite{SM} for details on the statistical analysis) for the onset of primary creep during the first loading step of one experiment, tough associated to a large $E_0 \sim 3 \times 10^4$ (arb. units).

By obtaining a maximum likehood estimate for the exponent $\beta$ we see (Fig.~\ref{fig:fig3}b) that its value does not change between experiments or between different creep steps. The observed value of $\beta = 1.47 \pm 0.04$ 
is very close to the mean-field value of $\beta=3/2$.\\

{\em Summary -} We studied primary creep of an emblematic heterogeneous material -- concrete -- under uniaxial compression.
We observed a power-law (Omori-type) decrease in the strain rate and a similar behavior in the number of acoustic events, supporting the viewpoint of microcracking as the mechanism of creep deformation. This brittle creep mechanism, different from the microscopic mechanisms generally invoked to explain concrete creep, is further confirmed by our results of ultrasonic monitoring that show a similar evolution of the attenuation of elastic waves within the material as creep proceeds.
It results from the interplay between thermal activation and elastic interactions leading to damage avalanches. Unlike what is observed for damage under athermal monotonic loading \cite{vu2019compressive,vu2020asymmetric}, the distribution of acoustic emission energies during primary creep follows a pure power-law with an exponent close to the mean-field value even at low stress and quite far from failure. Interestingly, this is reminiscent of the postseismic slip phase following large earthquakes, which is characterized by a power-law decrease of the slip velocity, $v\sim 1/(t+c)^p$ \cite{periollat2022transient}, as well as a similar decay of the aftershock triggering rate, the celebrated Omori's law~\cite{omori1895after}, while the seismic moments of these aftershocks are distributed according to a non-truncated power law -- the Gutenberg--Richter law \cite{kagan2013earthquakes}.

However, the main outcome of this work is the discovery of loading history effects leading to an unexpected slowing down of primary creep dynamics after the material has been previously damaged during former creep steps at lower stresses. From a progressive damage model in which thermal activation of local damage events is introduced using a KMC algorithm, we argue that this results from an aging-under-stress phenomenon. In amorphous materials such as glasses, aging characterizes the process by which a glass, below its transition temperature $T_g$, tries (very) slowly to reach thermal equilibrium by moving towards more stable energy wells \cite{arceri2022statistical}. This aging therefore modifies the excitation spectrum, increasing, upon mechanically loading the amorphous material, its strength and brittleness~\cite{ozawa2018random}.
In our case, the material ages during primary creep, i.e. under stress, assisted by thermal activation (as illustrated by the expression of the Arrhenius term $V_a\Delta\sigma_i/k_BT$ in the model). This depletes the easy-to-damage sites (the ones with the smallest stress gaps), making the material less prone to creep when subsequently loaded under a \emph{larger} stress. Upon increasing further the applied stress, one reaches finally a point at which this aging effect is counterbalanced by an increasing softening of the material and of the role of elastic interactions leading to a sustained damage avalanche activity and larger creep rates. For a given primary creep step, this aging is fast at the onset of loading but slows down through time as creep proceeds (Fig.~\ref{fig:fig2}), much like classical aging of glasses~\cite{arceri2022statistical}.

Besides their importance for the fundamental understanding of creep in disordered brittle materials, these history effects have practical consequences in material testing that cannot be underestimated, e.g. when using stress-stepping experiments to empirically determine a creep law. Neglecting the aging-under-stress can falsely make the material seem more creep-resistant than it actually is, which is a serious problem for structural materials.

Beyond brittle creep, one might expect similar aging-under-stress phenomena in other contexts, as already hinted by Cottrell for the creep of metals \cite{cottrell1952time}. In that case, creep results from dislocation motion, and this aging would result from the exhaustion of easy-to-deform sites during primary creep, resulting in a sort of statistical strain-hardening. An extension to other materials, such as amorphous media, would be worth exploring as well.\\

\begin{acknowledgments}
{\em Acknowledgments -} We thank Chi-Cong Vu for the sample preparation, Phil\'{e}mon Peltier for the implementation of the KMC algorithm, and additionally thank Mikko Alava and Carmen Miguel for useful discussions.

ISTerre is part of Labex OSUG@2020.
T.M. acknowledges funding from the European Union Horizon 2020 research and innovation programme under grant agreement No 857470 and from European Regional Development Fund via Foundation for Polish Science International Research Agenda PLUS programme grant No MAB PLUS/2018/8.
Aalto Science IT project is acknowledged for computational resources.
\end{acknowledgments}


%

\end{document}


\preprint{APS/123-QED}

\title{Supplementary material: History effects in the creep of a disordered brittle material}

\author{Tero M\"{a}kinen}
 \email{Corresponding author\\tero.j.makinen@aalto.fi}
\affiliation{%
 Department of Applied Physics, Aalto University,\\
 P.O.  Box 11100, 00076 Aalto, Espoo, Finland
}%
\affiliation{NOMATEN Centre of Excellence, National Centre for Nuclear Research, ul. A. Soltana 7, 05-400  Otwock-\'{S}wierk, Poland}
\author{J\'er\^ome Weiss}
\affiliation{%
 Univ. Grenoble Alpes, CNRS, ISTerre, 38000 Grenoble, France
}%
\author{David Amitrano}
\affiliation{%
 Univ. Grenoble Alpes, CNRS, ISTerre, 38000 Grenoble, France
}%
\author{Philippe Roux}%
\affiliation{%
 Univ. Grenoble Alpes, CNRS, ISTerre, 38000 Grenoble, France
}%


\date{\today}

\pacs{Valid PACS appear here}
\maketitle

\section*{Experimental setup} 

The initial loading ramp (as well as the stress increases between stress steps) was performed at a constant force rate of 100~N/s, corresponding to a stress rate of roughly 0.1~MPa/s. The load and axial displacements of the piston were monitored at a 5~Hz frequency.
The sample axial displacement was determined by subtracting the known elastic deformation of the loading frame from the measured axial displacement.

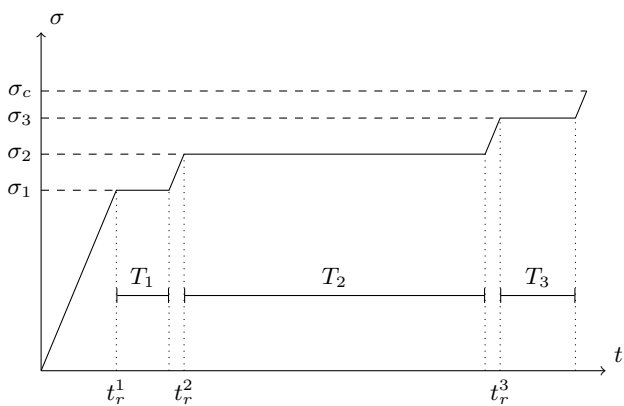
\begin{figure}[th!]
\begin{tikzpicture}
\def\slopex{1.0};
\def\slopey{2.4};
\def\creepone{0.7};
\def\creeptwo{4.0};
\def\creepthree{1.0};
\def\durHeight{1.0};

\draw[->] (0,0) -- (0, 4.5);
\draw[->] (0,0) -- (7.5, 0);
\node (stress) [anchor=south west] at (0, 4.5) {$\sigma$};
\node (time) [anchor=south west] at (7.5, 0) {$t$};

\draw (0,0) -- (\slopex, \slopey) -- (\slopex+\creepone, \slopey) -- (1.2*\slopex+\creepone, 1.2*\slopey) -- (1.2*\slopex+\creepone+\creeptwo, 1.2*\slopey) -- (1.4*\slopex+\creepone+\creeptwo, 1.4*\slopey) -- (1.4*\slopex+\creepone+\creeptwo+\creepthree, 1.4*\slopey) -- (1.55*\slopex+\creepone+\creeptwo+\creepthree, 1.55*\slopey);

\node (stress1) [anchor=east] at (0, \slopey) {$\sigma_1$};
\node (stress2) [anchor=east] at (0, 1.2*\slopey) {$\sigma_2$};
\node (stress3) [anchor=east] at (0, 1.4*\slopey) {$\sigma_3$};
\node (stressf) [anchor=east] at (0, 1.55*\slopey) {$\sigma_c$};
\draw[dashed] (0, \slopey) -- (\slopex, \slopey);
\draw[dashed] (0, 1.2*\slopey) -- (1.2*\slopex+\creepone, 1.2*\slopey);
\draw[dashed] (0, 1.4*\slopey) -- (1.4*\slopex+\creepone+\creeptwo, 1.4*\slopey);
\draw[dashed] (0, 1.55*\slopey) -- (1.55*\slopex+\creepone+\creeptwo+\creepthree, 1.55*\slopey);

\draw[dotted] (\slopex, 0) -- (\slopex, \slopey);
\draw[dotted] (\slopex+\creepone, 0) -- (\slopex+\creepone, \slopey);
\draw[dotted] (1.2*\slopex+\creepone, 0) -- (1.2*\slopex+\creepone, 1.2*\slopey);
\draw[dotted] (1.2*\slopex+\creepone+\creeptwo, 0) -- (1.2*\slopex+\creepone+\creeptwo, 1.2*\slopey);
\draw[dotted] (1.4*\slopex+\creepone+\creeptwo, 0) -- (1.4*\slopex+\creepone+\creeptwo, 1.4*\slopey);
\draw[dotted] (1.4*\slopex+\creepone+\creeptwo+\creepthree, 0) -- (1.4*\slopex+\creepone+\creeptwo+\creepthree, 1.4*\slopey);
\node (tr1) [anchor=north] at (\slopex, 0) {$t_r^1$};
\node (tr2) [anchor=north] at (1.2*\slopex+\creepone, 0) {$t_r^2$};
\node (tr3) [anchor=north] at (1.4*\slopex+\creepone+\creeptwo, 0) {$t_r^3$};
\draw [|-|] (\slopex,\durHeight) -- (\slopex+\creepone,\durHeight) node [midway, above] {$T_1$};
\draw [|-|] (1.2*\slopex+\creepone,\durHeight) -- (1.2*\slopex+\creepone+\creeptwo,\durHeight) node [midway, above] {$T_2$};
\draw [|-|] (1.4*\slopex+\creepone+\creeptwo,\durHeight) -- (1.4*\slopex+\creepone+\creeptwo+\creepthree,\durHeight) node [midway, above] {$T_3$};

\end{tikzpicture}
\caption{\label{fig:loading} A schematic picture of the loading protocol used. After an initial ramp of constant stress rate (ending at time $t_r^1$) the stress is kept constant at $\sigma_1$ for a duration $T_1$. The stress is then increased with the same stress rate to $\sigma_2$ for a duration of $T_2$ and so forth. The test ends when the sample fails either during a creep stage (stress step) or during one of the ramps (at the stress $\sigma_c$).}
\end{figure}

The testing protocol comprises of an initial loading ramp, static loading at constant stress $\sigma_1$ for a duration $T_1$, another loading ramp to $\sigma_2$, constant stress $\sigma_2$ for a duration $T_2$ and so forth until sample failure. A schematic representation of the loading protocol can be seen in Fig.~\ref{fig:loading}.\\

\begin{figure}[th!]
\begin{tikzpicture}
    \def\circRadius{2.0};
    \def\circumferenceYPos{2.7};
    \def\depthXPos{2.3};
    \def\widthYPos{2.5};
    \def\widthCut{0.3};
    \def\heightCut{0.5};
    
    \draw[dashed] (0,0) circle (\circRadius);
    
    \draw[<->] (-\circRadius,\circumferenceYPos) -- (\circRadius,\circumferenceYPos) node[above, midway] {$r$};
    \draw[<->] (\widthCut-\circRadius,\widthYPos) -- (\circRadius-\widthCut,\widthYPos) node[below, midway] {$W_1$};
    \draw[<->] (\depthXPos,\circRadius) -- (\depthXPos,\heightCut-\circRadius) node[right, midway] {$W_2$};
    
    \draw (\widthCut-\circRadius,{sqrt(2*\widthCut*\circRadius-\widthCut*\widthCut)}) -- (\widthCut-\circRadius,{-sqrt(2*\widthCut*\circRadius-\widthCut*\widthCut)});
    \draw (\circRadius-\widthCut,{sqrt(2*\widthCut*\circRadius-\widthCut*\widthCut)}) -- (\circRadius-\widthCut,{-sqrt(2*\widthCut*\circRadius-\widthCut*\widthCut)});
    \draw ({sqrt(2*\heightCut*\circRadius-\heightCut*\heightCut)},\heightCut-\circRadius) -- ({-sqrt(2*\heightCut*\circRadius-\heightCut*\heightCut)},\heightCut-\circRadius);
    
    \draw (\circRadius-\widthCut,{sqrt(2*\widthCut*\circRadius-\widthCut*\widthCut)}) arc[radius=\circRadius, start angle=atan(sqrt(2*\widthCut*\circRadius-\widthCut*\widthCut)/(\circRadius-\widthCut)), end angle=180-atan(sqrt(2*\widthCut*\circRadius-\widthCut*\widthCut)/(\circRadius-\widthCut))];
    \draw ({-sqrt(2*\heightCut*\circRadius-\heightCut*\heightCut)},\heightCut-\circRadius) arc[radius=\circRadius, start angle=180-atan((\heightCut-\circRadius)/sqrt(2*\heightCut*\circRadius-\heightCut*\heightCut)), end angle=180+atan(sqrt(2*\widthCut*\circRadius-\widthCut*\widthCut)/(\circRadius-\widthCut))];
    \draw ({sqrt(2*\heightCut*\circRadius-\heightCut*\heightCut)},\heightCut-\circRadius) arc[radius=\circRadius, start angle=atan((\heightCut-\circRadius)/sqrt(2*\heightCut*\circRadius-\heightCut*\heightCut)), end angle=-atan(sqrt(2*\widthCut*\circRadius-\widthCut*\widthCut)/(\circRadius-\widthCut))];

\node[inner sep=0, anchor=north] (sampleImg) at (0, -2.2) {\includegraphics[width=\columnwidth]{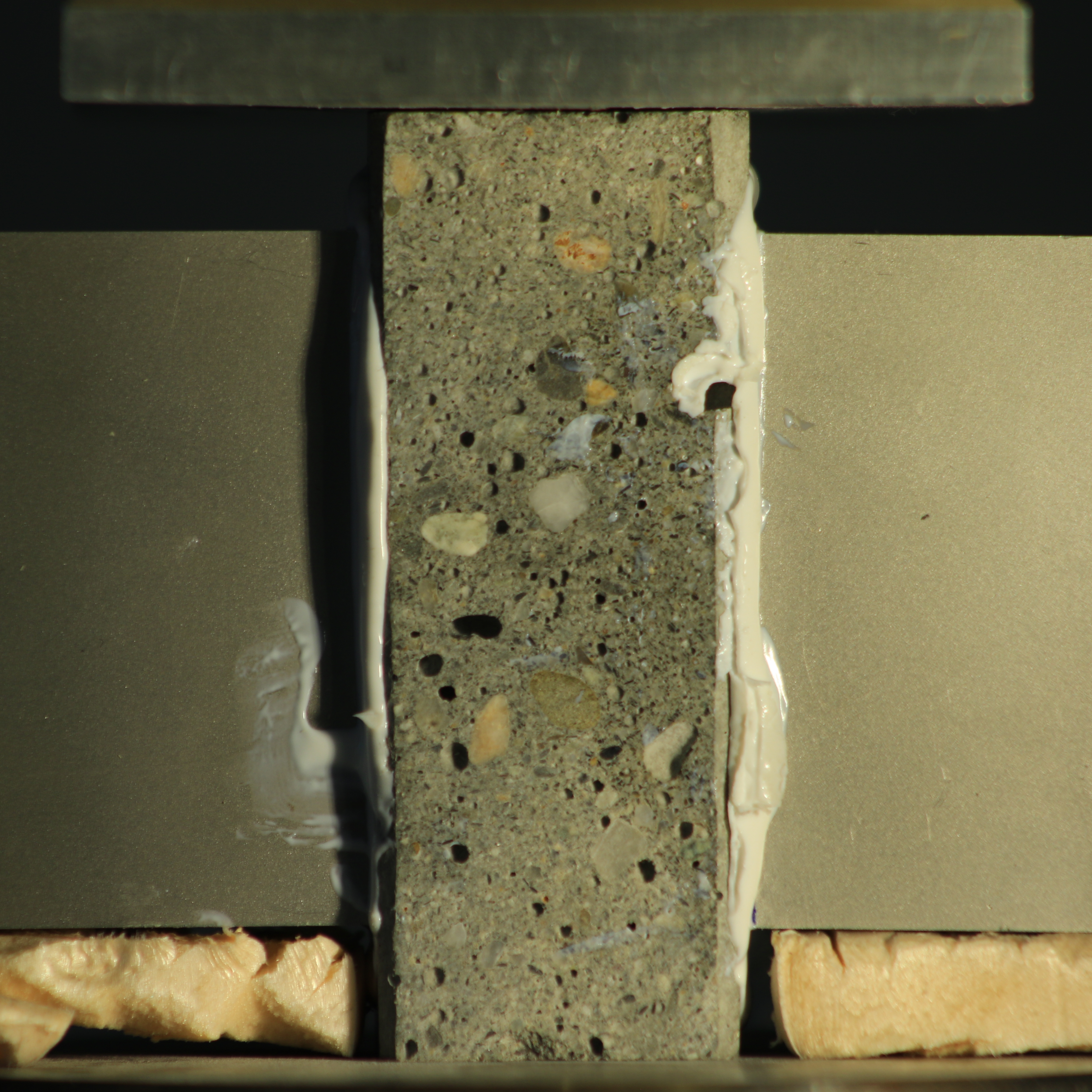}};
\node[inner sep=0, anchor=south west] (aLabel) at (-0.5\columnwidth, 2.7) {\textbf{a})};
\node[text=white, inner sep=0, anchor=south west] (bLabel) at (-0.49\columnwidth, -2.7) {\textbf{b})};
\draw[<->, white] (-2,-3.1) -- (-2,-10.6) node[left, midway] {$h$};
\node[text=white, inner sep=0, anchor=center] (piston) at (0, -2.6) {compression piston};
\node[text=white, inner sep=0, anchor=center, rotate=90] (leftArray) at (-0.35\columnwidth, -6.7) {ultrasonic array};
\node[text=white, inner sep=0, anchor=center, rotate=90] (rightArray) at (0.35\columnwidth, -6.7) {ultrasonic array};
\filldraw [fill=black, draw=none] (3, -10.2) rectangle (3+0.9352, -10.1) node[above, midway] {1 cm};
\end{tikzpicture}
\caption{\label{fig:setup} a) The geometry of the sample cross-section showing the initially cylindrical samples (radius $r$) and the three flattened sides.
b) An image of the sample in the compression setup, showing the attachment of the ultrasonic arrays, as well as the heterogeneity of the sample microstructure. The height of the sample is denoted in the image as $h$.}
\end{figure}

The samples used were cylindrical pieces of dry concrete with a radius of 40~mm and height of 80.5~$\pm$~0.5~mm. Three sides of the samples were cut flat for imaging purposes (see Fig.~\ref{fig:setup}a for a schematic representation) but the results of this imaging are not included in this manuscript (see Fig.~\ref{fig:setup}b for an image of the sample). The loading direction during compression was perpendicular to the direction of casting. The two types of concrete used correspond to aggregate sizes of roughly 2~mm (M-concrete) and 3.5~mm (C-concrete).
The sample preparation and detailed microstructural analysis of the samples have been detailed in Ref.~\cite{vu2018revisiting}.\\

The acoustic emission was captured by coupling two piezoelectric AE sensors (frequency bandwidth of 20-1200 kHz) directly to the side of the sample using a silicon paste. The resulting signals were preamplified by 40 dB and a standard thresholding procedure for AE event detection was used with a 30 dB amplitude threshold. The event catalog consists of the occurrence times of the events and their energies $E_{AE}$.\\

In addition, two source-receiver ultrasonic arrays face each other on both sides of the sample (Fig.~\ref{fig:setup}b). The arrays are composed of 64 transducers centered at 1~MHz with a 75 percent bandwidth. The ultrasonic signal transmitted by each piezo-transducer source is a broadband pulse of $1$ $\mu$s at the central frequency of the transducer. On both arrays, the transducer dimensions are 0.75~mm along the vertical axis (that corresponds to half of the central wavelength), and 12~mm along the transverse axis. This feature naturally creates a collimated beam, which avoids side echoes from the sample boundaries.
The received signals spread over $~10$ $\mu$s after the direct arrival at $\simeq 9$ $\mu$s due to weak scattering associated to the sample heterogeneity. To solely focus on the amplitude variation of the ballistic arrival, a Double Beamforming algorithm (DBF) is applied on 9-element subarrays centered on source-receiver pairs from both sides \cite{tudisco2015timelapse}.
The DBF algorithm presents two advantages. First, it allows us to clean out the scattered wavefield with the selection of the acoustic beam of maximum intensity for each source-receiver pair \cite{iturbe2009travel}. Second, it significantly improves the signal-to-noise ratio of the ballistic arrival when the elastic attenuation increases close to failure \cite{nicolas2008double}.

\section*{The simulation model}

The basic framework of the model is a well-studied progressive damage model~\cite{amitrano1999diffuse, girard2010failure, girard2012damage}, which utilizes a 2D plane strain Finite Element Method with a triangular mesh and adds the damage through a decrease in the elastic modulus
\begin{equation}
    E = (1-d) E_0
\end{equation}
where $E_0$ is the initial elastic modulus and $d$ the damage parameter. Each time an element is damaged (the criterion for this is explained below) the modulus is decreased by 10 \%.\\

The time-dependent part is implemented using the kinetic Monte Carlo (KMC) approach which works in two basic steps. First a site is picked randomly, with a probability proportional to the
jump rate which follows an activated scaling
\begin{equation}
    \nu_i = \nu_0 \exp\left( - \frac{E_i}{k_B T} \right)
\end{equation}
where the index $i$ corresponds to the site in the simulated sample, $\nu_0$ is the constant attempt frequency, $E$ the activation energy, $k_B$ the Boltzmann constant and $T$ the absolute temperature. The activation energy is determined as 
$E_i = V \Delta \sigma_i$
where $V$ is a (constant) activation volume and $\Delta \sigma$ is the size of the stress step in the Coulomb failure criterion at the current time.

After the site is picked, it is damaged according to the original formulation of the model. The second step of the KMC algorithm is then a random choice of the timestep from a distribution with the probability density function
\begin{equation}
    p(\Delta t) = \frac{1}{\Delta t_0} \exp\left( - \frac{\Delta t}{\Delta t_0} \right) \mathrm{.}
\end{equation}
where $\Delta t_0$ is the reciprocal of the sum of the jump rates $\sum_i \nu_i$.
The simulation time used is moved forward by this randomly chosen $\Delta t$ and the cycle starts again from the first step.\\

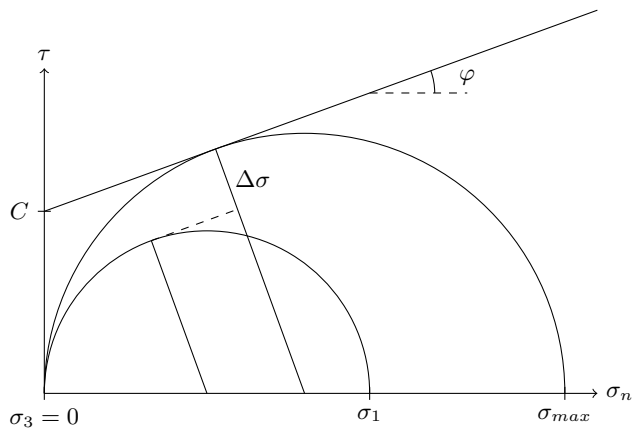
\begin{figure}
    \centering
    \begin{tikzpicture}

\def\axheight{0.5\columnwidth};
\def\axwidth{0.85\columnwidth};
\def\sigmaMax{0.8\columnwidth};
\def\sigmaOne{0.5\columnwidth};
\def\tickLength{0.01\columnwidth};
\def\theta{110}; 
\def\angleX{0.5\columnwidth};
\def\angleR{0.1\columnwidth};

\draw[->] (0, 0) -- (0, \axheight) node (tau) [above] {$\tau$};
\draw[->] (0, 0) -- (\axwidth, 0) node (sigmaNorm) [right] {$\sigma_n$};

\draw (\sigmaMax,0) arc(0:180:0.5*\sigmaMax);
\draw (\sigmaOne,0) arc(0:180:0.5*\sigmaOne);
\draw (\sigmaMax/2,0) -- ({\sigmaMax/2*(1+cos(\theta))},{\sigmaMax/2*sin(\theta)});
\draw (\sigmaOne/2,0) -- ({\sigmaOne/2*(1+cos(\theta))},{\sigmaOne/2*sin(\theta)});

\draw (0, {\sigmaMax/2*cot(\theta/2)}) -- (\axwidth, {\sigmaMax/2*cot(\theta/2)-cot(\theta)*\axwidth});
\draw [dashed] ({\sigmaOne/2*(1+cos(\theta))},{\sigmaOne/2*sin(\theta)}) -- ({(\sigmaMax/2+\sigmaOne/2*cot(\theta/2)*cot(\theta))/(1+(cot(\theta))^2)},{\sigmaOne/2*sin(\theta)+(\sigmaOne-\sigmaMax)/2*sin(\theta)*cos(\theta)});

\draw (-\tickLength, {\sigmaMax/2*cot(\theta/2)}) node (c) [left] {$C$} -- (\tickLength, {\sigmaMax/2*cot(\theta/2)});
\draw (0, -\tickLength) node (sigmaThree) [below] {$\sigma_3 = 0$} -- (0, \tickLength);
\draw (\sigmaOne, -\tickLength) node (sigmaOne) [below] {$\sigma_1$} -- (\sigmaOne, \tickLength);
\draw (\sigmaMax, -\tickLength) node (sigmaMax) [below] {$\sigma_{max}$} -- (\sigmaMax, \tickLength);
\draw ({(cos(\theta/2))^2*(\sigmaMax+(\sigmaOne-\sigmaMax)/2*cos(\theta))}, {(\sigmaOne+\sigmaMax+(\sigmaOne-\sigmaMax)*cos(\theta))*sin(\theta)/4}) node (deltaSigma) [right] {$\Delta \sigma$};

\draw [dashed] (\angleX, {\sigmaMax/2*cot(\theta/2)-\angleX*cot(\theta)}) -- (\angleX+1.5*\angleR, {\sigmaMax/2*cot(\theta/2)-\angleX*cot(\theta)}) node (phi) [above] {$\varphi$};
\draw (\angleX+\angleR,{\sigmaMax/2*cot(\theta/2)-\angleX*cot(\theta)}) arc(0:{\theta-90}:\angleR);

\end{tikzpicture}
    \caption{An illustration of the Coulomb failure criterion used, represented in the Mohr plane. The stress gap $\Delta \sigma$ is defined as the distance from the current stress state (circle defined by the applied normal stress $\sigma_1$) to the failure envelope (defined by the line with a slope $\tan \varphi$ and $y$-intercept $C$).}
    \label{fig:mohrCoulomb}
\end{figure}

The stress step used in the model is determined from the classical Coulomb failure criterion.
The critical normal stress $\sigma_{max}$ is determined by the envelope
\begin{equation}
    \tau = C_i + \sigma_n \tan \varphi
\end{equation}
where $\tau$ is the shear strength, $C_i$ the cohesion value of the site, $\sigma_n$ the normal stress, and $\varphi$ the angle of internal friction. The cohesion values for each site are picked from a quenched uniform distribution spanning from 0 to $C_{max}$.

The stress gap $\Delta \sigma$ is then the distance from the current state to the athermal failure state (see Fig.~\ref{fig:mohrCoulomb} for a graphical explanation). Here only normal stress is applied and therefore $\sigma_n = \sigma_1$ and the stress gap is
\begin{equation}
    \Delta \sigma 
    = C \cos \varphi  + \frac{\sigma_1+\sigma_3}{2} \sin \varphi - \frac{\sigma_1-\sigma_3}{2}
    \mathrm{.}
\end{equation}
The values of the model parameters chosen to mimic the macroscopic behavior observed experimentally can be seen in Table~\ref{tab:tableParam}.

\begin{table}[h]
\caption{\label{tab:tableParam}%
Parameter values for the model}
\begin{ruledtabular}
\begin{tabular}{llll}
$E_0$ [GPa] &
$d$ &
$\nu_0$ [s${}^{-1}$] & 
$T$ (K) \\
$21$ & $0.1$ & $1 \cdot 10^{13}$& $ 300 $ \\
\colrule
$V$ [m${}^3$] &
$\varphi$ [°]&
$C_{max}$ [MPa] &
 \\
$11\cdot 10^{-27}$ & $35$ & $31$ & \\
\end{tabular}
\end{ruledtabular}
\end{table}

\section*{Evolution of the $p$-exponent values}

\begin{figure}[h!]
    \centering
    \includegraphics[width=\columnwidth]{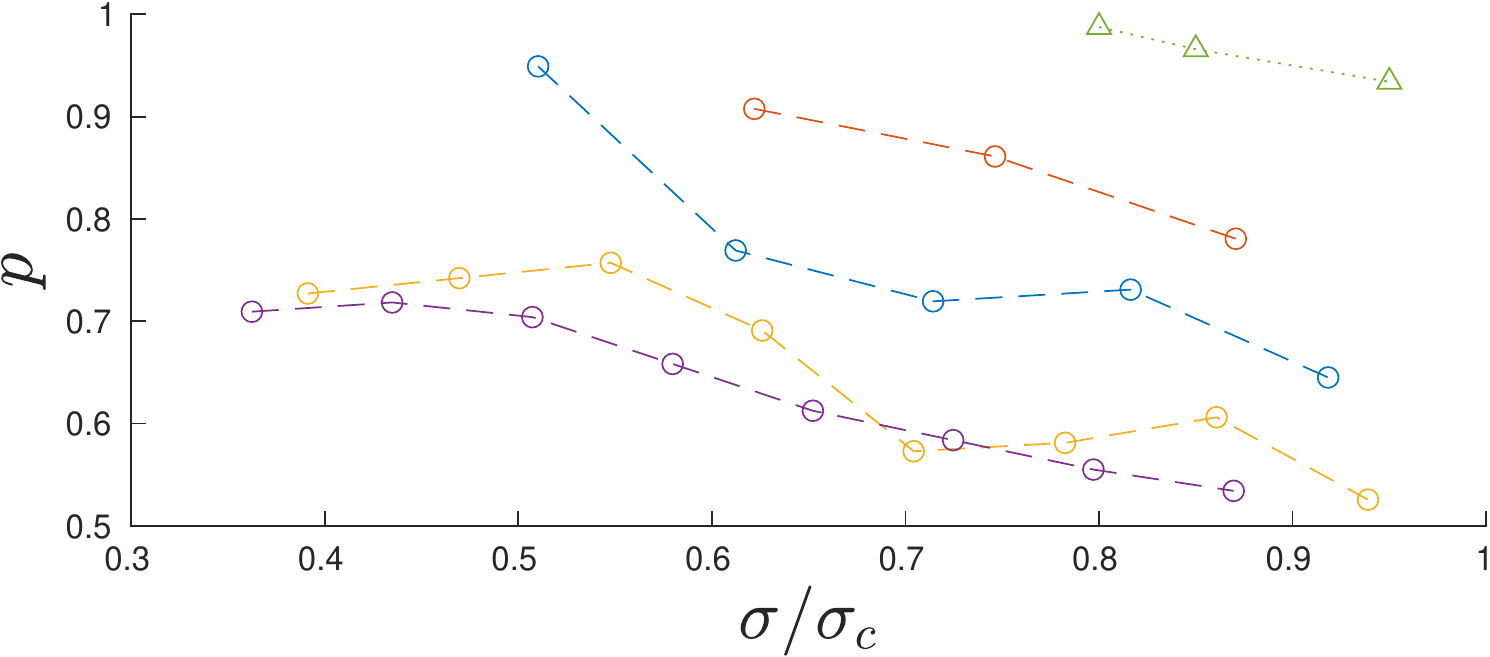}
    \caption{The evolution of the $p$ exponent in the Eq.~1 of the main paper as a function of the scaled applied stress. The triangle symbols correspond to the simulations and the circles to the experiments.}
    \label{fig:figP}
\end{figure}

In the main text we state that fitting the strain rate relation (Eq.~1 in the main text) to the observed strain curves (Fig.~1a of the main text) shows that the exponent $p$ decreases with increasing stress. This effect is shown in Fig.~\ref{fig:figP} where $p$ is plotted as a function of the scaled applied stress $\sigma/\sigma_c$. The decrease is clear in all the experiments (plotted with circles) although the initial value of the exponent varies from around 0.7 to almost unity. The same behavior is seen in the simulations (plotted with triangles) where the initial value of $p$ is very close to unity and decreases with increasing stress.

\section*{Statistical analysis of the acoustic emission} 

From the dataset of $N$ events with energies $E_{AE}^i$ (where $i \in [1, N]$) the parameters of the event energy distribution are estimated using the maximum likelihood method \cite{baro2012analysis} where one computes the (logarithm) of the likelihood function $\mathcal{L}$ and finds the parameters which maximize this function.\\

The natural choices for the distributions we see are a power-law (PL)
\begin{equation}
    p(E_{AE}) = \frac{E_{AE}^{-\beta}}{\zeta_{PL}}
\end{equation}
where $\zeta_{PL}$ is the normalization factor
and the truncated power-law (TPL) or power-law with an exponential cutoff
\begin{equation}
    p(E_{AE}) = \frac{E_{AE}^{-\beta} \exp(-E_{AE}/E_0)}{\zeta_{TPL}}
\end{equation}
which give the log-likelihoods
\begin{equation}
    \ln \mathcal{L}_{PL} = -\beta \sum_{i=1}^N \ln E_{AE}^i - N \ln \zeta_{PL}
\end{equation}
and
\begin{equation}
    \ln \mathcal{L}_{TPL} = -\beta \sum_{i=1}^N \ln E_{AE}^i + \sum_{i=1}^N \frac{E_{AE}^i}{E_0} - N \ln \zeta_{TPL}
    \mathrm{.}
\end{equation}

\begin{figure}[b!]
    \centering
    \includegraphics[width=\columnwidth]{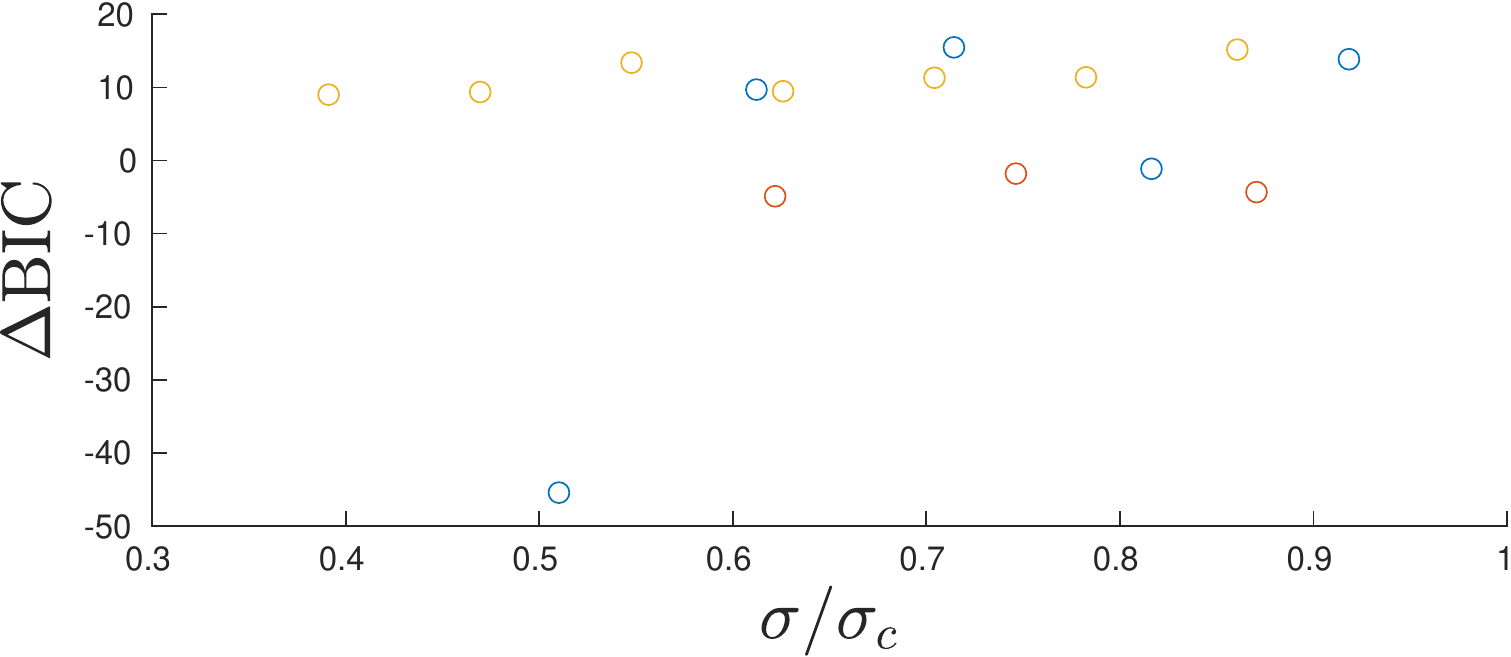}
    \caption{The evolution of the difference of the BICs for the power-law and truncated power-law models (Eq.~\ref{eq:deltaBic}) in the experiments as a function of the scaled applied stress, showing that the truncated power-law is significantly favored in only one of the stress steps.}
    \label{fig:figDeltaBic}
\end{figure}

As the two models have a different number of parameters, a direct comparison of the likelihoods is not the best way to compare their goodness-of-fit.
Instead for comparison one can compute the Bayesian Information Criterion (BIC) \cite{bell2013convergence, vasseur2015heterogeneity}
\begin{equation}
    \mathrm{BIC} = - 2 \ln \mathcal{L} + 2 N_p \ln N
\end{equation}
where $N_p$ is the number of parameters. The more illustrative quantity is the difference of the BICs for the power-law and truncated power-law models
\begin{equation} \label{eq:deltaBic}
    \Delta \mathrm{BIC} = 2 \left(\ln \mathcal{L}_{PL} - \ln \mathcal{L}_{TPL} - \ln N \right) 
\end{equation}
which is negative if the truncated power-law is favored, positive if the pure power-law is favored and around zero if the models give equally good results.

Plotting the $\Delta \mathrm{BIC}$ as a function of the scaled applied stress (Fig.~\ref{fig:figDeltaBic}) shows that it is fairly close to zero for almost every stress step in the experiments. This is expected, as with a high cutoff value the power-law and truncated power-law should yield similarly good fits. However for a single stress step (the first step in the experiment presented in the main text) the truncated power-law is clearly preferred, but also here the cutoff value is fairly high (as stated in the main text).

\section*{Additional experiments}

\begin{figure}[h!]
    \centering
    \includegraphics[width=\columnwidth]{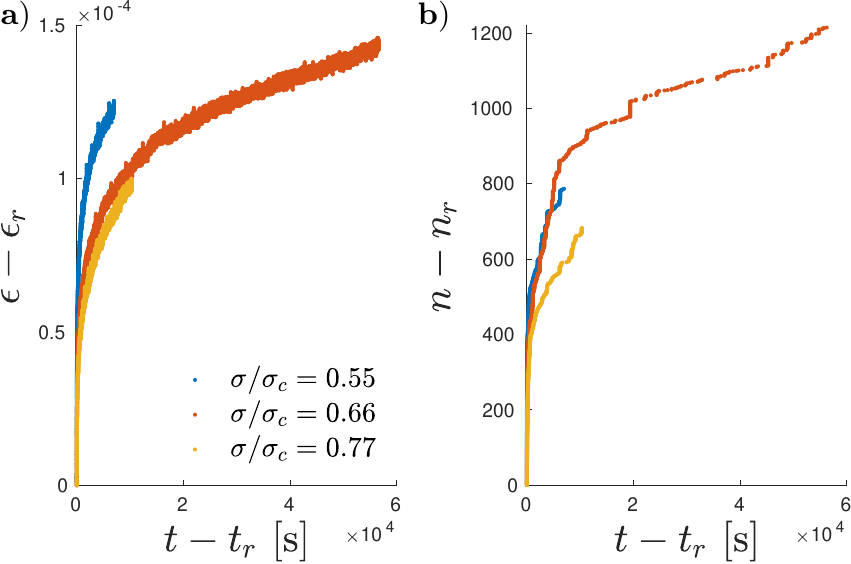}
    \caption{a) An additional experiment with C-concrete showing that the behavior with this microstructure is very similar to the M-concrete: after an initial stress step the creep is slower and as the second step is fairly long, the creep is even slower in the third step. b) The number of acoustic events shows behavior which is very similar to the time evolution of the strain.}
    \label{fig:fig1_G1-3}
\end{figure}

\begin{figure}[h!]
    \centering
    \includegraphics[width=\columnwidth]{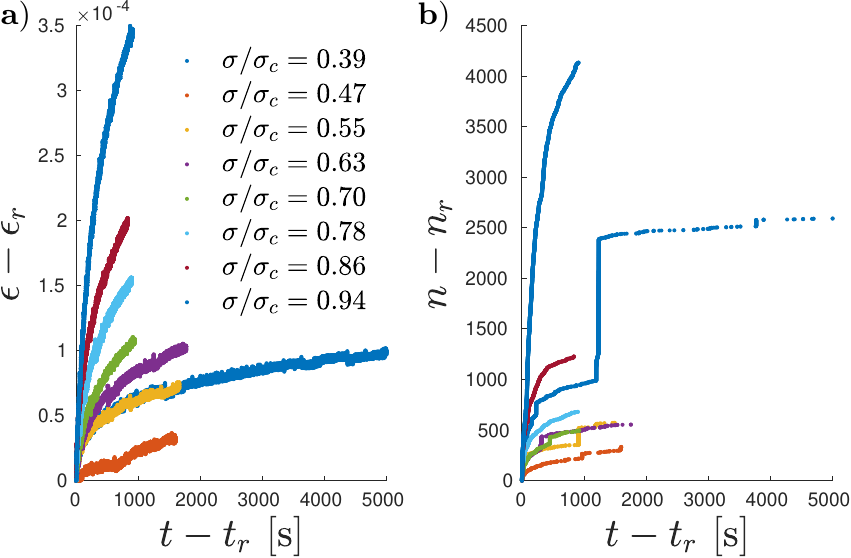}
    \caption{a) An additional experiment with M-concrete, showing the effect of an extremely long initial stress step (real duration 64036 s, plot truncated for clarity) on subsequent, short stress steps. After the extremely long initial step the next (fairly short) step shows very slow creep and in the third step the creep rate is only roughly equal to the first one. b) Except for one sudden burst of events in the first stress step, the number of acoustic events roughly follows the behavior of the strain.}
    \label{fig:fig1_M4-1}
\end{figure}

\begin{figure}[h!]
    \centering
    \includegraphics[width=\columnwidth]{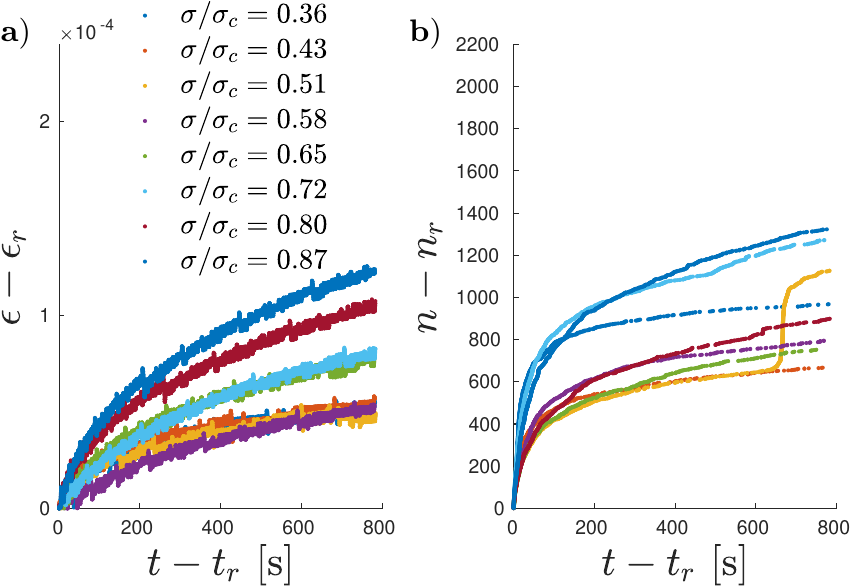}
    \caption{a) A reference experiment with M-concrete where all the stress steps are kept short (equal durations). The creep rates for the first three steps are roughly equal and after that start to increase. This behavior is not consistent with the power-law model $\dot{\epsilon} \propto \sigma^n$. b) Also here the behavior of the number of acoustic events roughly matches the behavior of strain. There is again a sudden burst of events close to the end of one stress step.}
    \label{fig:fig1_M-C1-4-25}
\end{figure}

For the sake of clarity only one representative experiment is presented in the main text, but several other experiments were performed, with varying durations of stress steps as well as microstructures. We have included the strain and event rate behaviors of these experiments in Figs.~\ref{fig:fig1_G1-3}--\ref{fig:fig1_M-C1-4-25}. History effects similar to those analyzed for the representative experiment detailed in the main text are observed in each case, for both the strain and the number of AE events.  


%